\documentclass[12pt]{article}
																			 																																																												\pdfoutput=1	
\usepackage{amsmath,amssymb}
\usepackage{graphicx}
\usepackage[pdftex]{hyperref}
\usepackage{epstopdf}
\DeclareGraphicsExtensions{.eps,.bmp,.wmf,.jpeg,.pdf}
\numberwithin{equation}{section}
\def\be{\begin{equation}}
\def\ee{\end{equation}}

\def\bea{\begin{eqnarray}}
\def\eea{\end{eqnarray}}
\title{Modified gravity with an exponential function of curvature}
\author{L. N. Granda\thanks{luis.granda@correounivalle.edu.co}\\ {\small\it Departamento de Fisica, Universidad del Valle}\\{\small\it A.A. 25360, Cali, Colombia}}
\date{}
\begin{document}
\bibliographystyle{alpha}
\maketitle

\begin{abstract}
\noindent 
The role of an exponential function of the scalar curvature in the modified gravity is analyzed. Two models are proposed. A toy model that complies with local and cosmological constraints and gives appropriate qualitative description of the cosmic evolution. This model contains a saddle matter-dominant critical point that can evolve towards a late time de Sitter attractor. Initial conditions have been proposed, showing that this toy model has an acceptable matter era and gives an approximate qualitative behavior of cosmic evolution. A second viable model, behaves very close to $\Lambda$CDM at early times and can satisfy local and cosmological constraints. It behaves as $R-2\Lambda$ at $R\rightarrow \infty$ and tends to zero at $R\rightarrow 0$, containing flat spacetime solution. The model gives viable cosmological trajectories that, as the first model, connect the matter dominated point with a late time de Sitter attractor. The cosmic evolution of the main density parameters in this model is consistent with current observations with an equation of state very close to $-1$.

%\begin{description}

%\item[PACS numbers]
%98.80.-k, 04.50.kd, 95.36+x
%\end{description}
\end{abstract}

\section{Introduction}
\noindent 
So far the most successful dark energy model is the cosmological constant (for review see \cite{copeland, sahnii, padmanabhan, sergeiod}), despite its main fine-tuning problem, that motivates the seek for alternative models of dynamical nature. Among these models, the modification of gravity that involves a general function $f(R)$, represents an appealing alternative that has been under intense study last years. The function $f(R)$ generalizes the Einstein-Hilbert Lagrangian by adding corrections that are non-linear functions of the curvature, subject to local (solar system) and cosmological (high redshifts) constraints that determine its viability (see \cite{sodintsov1, sotiriou, tsujikawa0, sodintsov1a, odin-nojiri, odin-oiko4} for reviews). These corrections may become relevant in a late universe and many types of modifications to the Einstein-Hilbert action have been proposed so far \cite{sodintsov1, capozziello, capozziello1, sodintsov, carroll, faraoni, dobado, anthoni, barrow1, nojiri,  elizalde, troisi, allemandi, koivisto, brevik, sodintsov2, nojiri1, nojiri2, olmo, hu}. Among the first and most studied corrections to the Einstein-Hilbert Lagrangian are the corrections of the form $R^n$, but it is well known that corrections with $n>1$ that are relevant at early times like in the case of $n=2$ leading to de Sitter expansion \cite{starobinsky}, are negligible small compared to $R$ at the present epoch and not suitable to explain the current accelerated expansion. Models with $n< 0$ contain instabilities that prevent them from having a matter dominated era \cite{dolgov, lucamendola, sodintsov2} and are also inconsistent with solar system tests. There are also models that attempt to unify early time inflation with late time acceleration \cite{nojiri5, nojiri6, nojiri7, nojiri8}. Modified gravity with arbitrary function of the 4-dimensional Gauss-Bonnet invariant has been introduced in  \cite{nojiriodintsov1, nojiriodintsov2, cognolaelizalde}.
Any viable model of modified gravity should pass not only the Solar-system tests, that are perhaps the more reliable and challenging, where the average density of matter is high compared with that of the universe, but also should satisfy the cosmological restrictions from high redshift observations. 
The so called chameleon mechanism is used to pass solar system tests. The purpose of this mechanism is to give a large enough mass to the scalar field (that appears after the conformal transformation in the metric to convert $f(R)$ to the Einstein frame) to avoid measurable corrections to the local gravity  phenomena \cite{hu, tsujikawa, brax}. A number of works have been devoted to $f(R)$ models that can satisfy both cosmological and local gravity constraints  \cite{hu, astarobinsky, appleby1, sergeid1, sergeid2, eelizalde}.  Exact cosmological solutions have been studied in \cite{bamba1, barrow, clifton, capozz, capozz1, capozz2, barrow2}. The theory of dynamical systems is a very useful tool to study models with highly non-linear field equations and have been widely applied in cosmology (see \cite{sebastian} for revision). The dynamical systems encode many important features of the models, where in the case of cosmology, the critical points and  their stability properties describe the different phases of evolution of the universe. Different cosmological aspects of modified gravity models, using dynamical system techniques, have been studied in   
\cite{amendola1, tsujikawa1, naureen, souza, tsujikawa0, gsaridakis2, boehmer1, kofinas, muller, boehmer, mirza, gsaridakis1, landim, bernardi, odin-oiko2, odin-oiko3, odin-oiko, odin-oiko1, oikonomoua}.\\
In the present paper we consider an exponential function of the curvature in modified gravity and study its cosmological consequences. It is shown that the two proposed models can satisfy both, local and large scale cosmological constraints. As the criterium to analyze the outcomes of the models we used the ($m,r$) diagram which shows that the models are viable and contain the matter era followed by a late time solution with accelerated expansion. The first model gives an approximate qualitative description of the cosmic evolution while the second second model is more realistic and behaves very close to the $\Lambda$CDM with disappearing cosmological constant at $R\rightarrow 0$. A simple modified gravity model with exponential gravity, that realize early and late time accelerated expansion, was proposed in \cite{elizalde1} and observational constraints on this model were studied in \cite{sergeisaez}. 
A more general model with exponential and logarithmic corrections was considered in \cite{sergeisaez1} and constant roll inflation with exponential modified gravity was studied in \cite{sergeioiko}.\\
This paper is organized as follows. In section 2 we present the general features of the $f(R)$ models, including the dynamical system and the relevant critical points for our study in terms of the $(r,m)$ parameters. In section 3 we present the models, showing the conditions for viability and its trajectories in the ($r,m$)-plane, and some numerical cases of cosmic evolution. In section 4 we present some discussion.
%%%%%%%%%%%%%%%%%%%%%%%%%%%%%%%%%%%%%%%%%%%%%%%%%%%%%%%%%%%%%%%%%%%%%%%%%%%%%%%%%%%%%%%%%%%%%%%%%%%%%%%%%%%%%%%%%%%%%%%%%%%%%
\section{Field equations}
Let us start with the following action for modified gravity
\be\label{eq1}
S=\int d^4x\sqrt{-g}\left[\frac{1}{2\kappa^2}f(R)+{\cal L}_m\right]
\ee
where $\kappa^2=8\pi G$ and ${\cal L}_m$ is the Lagrangian density for the matter component which satisfies the usual conservation equation.  
Variation with respect to the metric gives the equation of motion
\be\label{eq2}
f_{,R}(R)R_{\mu\nu}-\frac{1}{2}g_{\mu\nu}f(R)+\left(g_{\mu\nu}\Box-\nabla_{\mu}\nabla_{\nu}\right)f_{,R}(R)=\kappa^2 T_{\mu\nu}^{(m)}
\ee
where $T^{(m)}_{\mu\nu}$ is the matter energy-momentum tensor assumed as 
$$T^m_{\mu\nu}=\left(\rho+p\right)u^{\mu}u^{\nu}+pg_{\mu\nu}$$
and $f_{,R}\equiv \frac{df}{dR}$. The trace of eq. (\ref{eq2})  gives 
\be\label{eq3}
Rf_{,R}(R)-2f(R)+3\Box f_{,R}(R)=\kappa^2 T^{(m)}=\kappa^2\left(3p-\rho\right)
\ee
The time and spatial components of the Eq. (\ref{eq2}) are given by the following expressions
\be\label{eq2a}
3H^2f_{,R}=\frac{1}{2}\left(Rf_{,R}-f\right)-3H\dot{f}_{,R}+\kappa^2\rho
\ee
and
\be\label{eq2b}
-2\dot{H}f_{,R}=\ddot{f}_{,R}-H\dot{f}_{,R}+\kappa^2\left(\rho+p\right)
\ee
where dot represents derivative with respect to cosmic time. The field equation (\ref{eq2a}) can be written in more compact form by defining the effective energy density as follows
\be\label{effdensity}
H^2=\frac{\kappa^2}{3}\rho_{eff},
\ee
where
\be\label{effdensity1}
\rho_{eff}=\frac{1}{f_{,R}}\left[\frac{1}{2\kappa^2}\left(Rf_{,R}-f-6H\dot{f}_{,R}\right)+\rho\right]
\ee
The Eqs. (\ref{eq2a}) and (\ref{eq2b}) lead to the following effective equation of state (EoS)
\be\label{eos}
w_{eff}=-1-\frac{2\dot{H}}{3H^2}=-1+\frac{\ddot{f}_{,R}-H\dot{f}_{,R}+\kappa^2\left(\rho+p\right)}{\frac{1}{2}\left(Rf_{,R}-f\right)-3H\dot{f}_{,R}+\kappa^2\rho},
\ee
where $\rho$ and $p$ include both matter and radiation components, i.e. $\rho=\rho_m+\rho_r$ and $p=p_m+p_r$. In order to be viable, the function $f(R)$ must satisfy the observational evidence both at the local level and at cosmological distances. The first general restrictions can be summarized as follows. Firstly the condition $f_{,R}>0$ is necessary to avoid negative effective Newtonian coupling. On the other hand, the scalar particle associated with $f(R)$, dubbed scalaron with mass (in matter epoch or in the regime $M^2>>R$)
 \be\label{mass}
 M^2\simeq \frac{1}{3f_{,RR}},
 \ee 
requires $f_{,RR}>0$ in order to avoid ghosts and is also a condition of stability under perturbations. \\
%This mass $M$ defines a range of the force mediated by the scalaron which determines the Compton wavelength $\lambda_C=2\pi M^{-1}$. If $\ell$ is the typical size of a local gravitational system, then the local gravity constraints on $f(R)$ are satisfied whenever $\ell>>\lambda_C$ or $M\ell>>1$ \cite{lamendola1, olmo}. \\
To study the viability of modified gravity as cosmological model it is useful to consider the dynamical system with the following dimensionless variables that can be obtained from Eq. (\ref{eq2a})) \cite{amendola1, tsujikawa1, tsujikawa0} (in what follows we will use indistinctly $f_{,R}$ or $F=f_{,R}$)
\be\label{autonomous}
x=-\frac{\dot{F}}{HF},\;\; y=-\frac{f}{6H^2 F},\;\; z=\frac{R}{6H^2}=\frac{\dot{H}}{H^2}+2,\;\; w=\frac{\kappa^2\rho_r}{3H^2F},\;\;  \Omega_m=\frac{\kappa^2\rho_m}{3H^2F}
\ee
which yield the following dynamical system
\be\label{restriction}
x+y+z+w+\Omega_m=1
\ee
\be\label{aut1}
\frac{dx}{dN}=x^2-xz-3y-z+w-1
\ee
\be\label{aut2}
\frac{dy}{dN}=xy+\frac{xz}{m}-2y(z-2)
\ee
\be\label{aut3}
\frac{dz}{dN}=-\frac{xz}{m}-2z(z-2)
\ee
\be\label{aut4}
\frac{dw}{dN}=xw-2zw
\ee
where $N=\ln a$, $w=\Omega_r$ is the density parameter of the radiation component, and the following quantities help to understand the viability of $f(R)$ models
\be\label{mr}
m=\frac{Rf_{,RR}}{f_{,R}},\;\;\;\; r=-\frac{Rf_{,R}}{f}.
\ee
In terms of these variables the effective EoS (\ref{eos}) is written as
\be\label{weffaut}
w_{eff}=-\frac{1}{3}\left(2z-1\right), 
\ee
while the dark energy equation of state from (\ref{eq2a}) and (\ref{eq2b}) can be written as \cite{amendola1, tsujikawa1, tsujikawa0}
\be\label{wde}
w_{DE}=-\frac{1}{3}\frac{2z-1+(F/F_0)w}{1-(F/F_0)(1-x-y-z)},
\ee
where $F_0$ is the current value of $f_{,R}$.\\
The critical points of the above dynamical system, in absence of radiation ($w=0$), for the model (\ref{eq1}) can be written in terms of $m$ and there are three important fixed points  \cite{amendola1, tsujikawa1, tsujikawa0} that we will consider to analyze the viability of our model: the critical point that gives rise to scaling solutions including the matter dominated era given by
\be\label{critpoints}
P_S=(x_c,y_c,z_c)=\left(\frac{3m}{1+m},-\frac{1+4m}{2(1+m)^2},\frac{1+4m}{2(1+m)}\right),
\ee
 with the following main parameters 
\be\label{critpointsa}
\Omega_m=1-\frac{m(7+10m)}{2(1+m)^2},\;\; w_{eff}=-\frac{m}{1+m},
\ee
and eigenvalues 
\be\label{autpoints}
EV(P_S): \left(3(1+m'),\frac{-3m\pm\sqrt{m(256m^3+160m^2-31m-16)}}{4m(m+1)}\right),
\ee
where prime represents derivative with respect to $r$. 
And the other two stable fixed points that lead to de Sitter and accelerated solutions
\be\label{critpointa}
P_{deS}=(x_c,y_c,z_c)=(0,-1,2),\;\;\; \Omega_m=0,\;\;\; w_{eff}=-1
\ee
with eigenvalues
\be\label{autpointa}
EV(P_{deS}): \left(-3,-\frac{3}{2}\pm \frac{\sqrt{25-16/m(r=-2)}}{2}\right),
\ee
and
\be\label{critpointb}
P_C=(x_c,y_c,z_c)=\left(\frac{2(1-m)}{1+2m},\frac{1-4m}{m(1+2m)},-\frac{(1-4m)(1+m)}{m(1+2m)}\right),
\ee
with the main parameters
\be\label{critpointb1}
\Omega_m=0,\;\; w_{eff}=\frac{2-5m-6m^2}{3m(1+2m)},
\ee
and the corresponding eigenvalues
\be\label{autpointb}
EV(P_C): \left(-4+\frac{1}{m},\frac{2-3m-8m^2}{m(1+2m)},-\frac{2(m^2-1)(1+m')}{m(1+2m)}\right).
\ee
From the coordinates $y$ and $z$ for the points $P_S$ and $P_C$ it can be seen that they are connected by the line $m(r)=-1-r$, where the relation $r=z/y$ is used.\\
From (\ref{critpoints}) follows that the matter dominated point corresponds to $(r,m)$=$(-1,0)$. The existence of a viable saddle matter era requires $m(r\rightarrow -1)>0$  and $-1<dm/dr(r\rightarrow -1)\le 0$. This last condition implies that all the $m(r)$ trajectories must be between the lines $m=0$ and $m=-r-1$. 
In order to be viable, the trajectory of a given $f(R)$ model in the $(r,m)$ plane should be such that it contains the matter dominated point $P_M=(-1,0)$ and starting from $P_M$ intersects the line $r=-2$ in the region $0<m\le 1$ \cite{amendola1}. The $\Lambda$CDM model, for instance, connect the points $P_M=(-1,0)$ and $P_{dS}=(-2,0)$. There are also viable trajectories connecting the saddle matter point $P_M=P_S(m\rightarrow 0)$ with the curvature dominated point that leads to stable accelerated expansion $P_C$, whenever $m'>-1$. \\
%We can appreciate the implications of the local-systems constraints on $m$ as follows. From the expression (\ref{mass}) for $M$ (valid at high curvatures when $Rf_{,RR}<<1$ and $f_{,R}\rightarrow 0$) and (\ref{mr}) it follows 
%\be\label{local}
%M\simeq \left(\frac{R_s}{3m(R_s)f_{,R_s}}\right)^{1/2}
%\ee
%where $R_s$ is the curvature of the local system. Then, from the condition $M\ell>>1$ it follows that
%\be\label{local1}
%m(R_s)<<\frac{\ell^2 R_s}{f_{,R}}.
%\ee
%Assuming $f_{,R_s}\sim 1$ and using the fact that the curvature $R_s\propto \rho_s$ and $R_0\sim H_0^2\propto \rho_0$ (where the subscript $"0"$ is for the present values) we can write
%\be\label{local2}
%m(R_s)<<\frac{\rho_s}{\rho_0}\left(\frac{\ell}{H_0^{-1}}\right)^2.
%\ee
%For the solar system, for instance, $\rho_s\sim 10^{-24}$ $gr/cm^3$, $\ell\sim 1.5\times 10^{13} cm$ and taking $\rho_0\sim 10^{-29}$ $gr/cm^3$, $H_0^{-1}\sim 10^{28} cm$ we find $m(R_s)<< 10^{-25}$. So, in high-curvature regions the quantity must be extremely  small.
%%%%%%%%%%%%%%%%%%%%%%%%%%%%%%%%%%%%%%%%%%%%%%%%%%%%%%
\section{The models}
{\bf Model 1.}\\
Firstly we discuss a toy model that  satisfies all above discussed requirements, given by the following function
\be\label{model}
f(R)=R e^{-\left(\frac{\mu^2}{R}\right)^{\eta}},
\ee
where $\eta>0$. 
This function is well defined everywhere in the interval $0\le R<\infty$ and can be expanded as
\be\label{frexpansion}
f(R)=R\left(1-\left(\frac{\mu^2}{R}\right)^{\eta}+\frac{1}{2}\left(\frac{\mu^2}{R}\right)^{2\eta}-....\right)=R-\mu^{2\eta}R^{1-\eta}+\frac{1}{2}\mu^{4\eta}R^{1-2\eta}-...
\ee
as follows from above expression the correction to $R$ is encoded in a convergent series that may contain positive (finite number in the case $0<\eta<1$) and infinite number of negative powers of curvature. Power-law corrections (with finite number of terms) have been considered in various works to tackle the dark energy problem, but have been shown to lead to non-viable cosmologies \cite{dolgov, lucamendola, sodintsov2}. The property that the exponential function grows faster than any power leads to the fact that any derivative of $f(R)$, due to the exponential function as given in (\ref{model}), is always well defined both when $R\rightarrow 0$ and when $R\rightarrow \infty$ which is useful when analyzing certain types of singularities \cite{eelizalde}. Particularly since $f_{,R}\rightarrow 1$ and $f_{RR},\; f_{RRR},... \rightarrow 0$ at $R\rightarrow \infty$, then, in absence of matter contribution, $\rho_{eff}$ in Eq. (\ref{effdensity1}) for the model (\ref{model}) becomes finite on singular solutions with $R\rightarrow \infty$, avoiding in this way type I and type III singularities.\\

\noindent The coefficient of $R$ in (\ref{model}) has the limits
\be\label{model1}
\lim_{R\to 0}e^{-\left(\frac{\mu^2}{R}\right)^{\eta}} =0,\;\;\; \lim_{R\to \infty}e^{-\left(\frac{\mu^2}{R}\right)^{\eta}}=1.
\ee
The first limit allows the existence of flat spacetime solutions and the second facilitates the consistency with high redshift CMB observations. This function can also be written as
\be\label{model3}
f(R)=R+\tilde{f}(R),\;\;\; \tilde{f}(R)=R\left(e^{-\left(\frac{\mu^2}{R}\right)^{\eta}}-1\right),
\ee
where the correction $\tilde{f}(R)$ and its derivative satisfy the condition (given $\eta>0$) $\lim_{R\to \infty} \tilde{f}(R)/R=0$ and $\lim_{R\to \infty} \tilde{f'}(R)=0$, which are important to recover the General Relativity at early times to satisfy the restrictions from Big Bang nucleosynthesis and CMB, and at at high curvature regime for local system tests. Taking the derivatives of $f(R)$ we find
\be\label{stability1}
f_{,R}=e^{-\left(\frac{\mu^2}{R}\right)^{\eta}}\left(1+\frac{\eta\mu^{2\eta}}{R^{\eta}}\right)>0,
\ee
indicating that the model satisfies automatically the condition $f_{,R}>0$, necessary to avoid antigravity regime. The second derivative gives
\be\label{stability2}
f_{,RR}=e^{-\left(\frac{\mu^2}{R}\right)^{\eta}}\left(\frac{\eta(1-\eta)\mu^{2\eta}}{R^{\eta+1}}+\frac{\eta^2\mu^{4\eta}}{R^{2\eta+1}}\right),
\ee
which satisfies the condition $f_{,RR}>0$ whenever
\be\label{frrpositive}
1-\eta+\frac{\eta^2\mu^{2\eta}}{R^{\eta}}>0.
\ee
This last inequality is always satisfied for $0<\eta\le 1$, but given the fact that the last term is positive, it also allows $1\le \eta<1+\eta\mu^{2\eta}/R^{\eta}$. For $0<\eta<1$, the correction to the General Relativity from the model (\ref{model}) contains a finite number of positive powers of $R$, $R^{\gamma_i}$ ($0<\gamma_i<1$) and an infinite number of negative powers of R, while for $\eta>1$ the correction contains only negative powers of $R$, but in both cases the correction is a regular function at $R\rightarrow 0$ and $R\rightarrow \infty$.\\
To continue the analysis we use the parameters $m$ and $r$ for this model, which are given by
\be\label{mr}
 m=\frac{\eta\left(\frac{\mu^2}{R}\right)^{\eta}\left[1-\eta+\eta\left(\frac{\mu^2}{R}\right)^{\eta}\right]}{1+\eta\left(\frac{\mu^2}{R}\right)^{\eta}},\;\;\; r=-1-\eta\left(\frac{\mu^2}{R}\right)^{\eta},
 \ee
 which gives the following relationship 
 \be\label{mr1}
 m(r)=-\frac{(r+1)(r+\eta)}{r}.
 \ee
As follows from this expression the model contains the matter dominated point $P_M=(-1,0)$. According to (\ref{autpointa}), the de Sitter $P_{deS}$ point is stable if  $0<m(r=-2)\le 1$. Applying this condition on (\ref{mr1}) leads to $0\le\eta<2$ which after the interception with the condition $f_{,RR}>0$ leads to the allowed values for $\eta$ 
\be\label{eta}
0<\eta\le 1
\ee
On the other hand, as follows from the expression (\ref{mr}) for $r$, the physically allowed values of $r$ satisfy the inequality $r<-1$, which imply that $r$ can approach $-1$ only form the left, i.e. $r\rightarrow -1^{-}$. This implies at the same time, according to  (\ref{mr}) and (\ref{eta}), that  $m$ approaches $0$ only form positive values, i.e. $m\rightarrow 0^{+}$ and the point becomes saddle spiral provided that $m'(-1)\ge -1$.
The Fig. 1 shows the possible trajectories in the $(r,m)$ plane for the model (\ref{model})
\begin{figure}
\begin{center}
\includegraphics[scale=0.7]{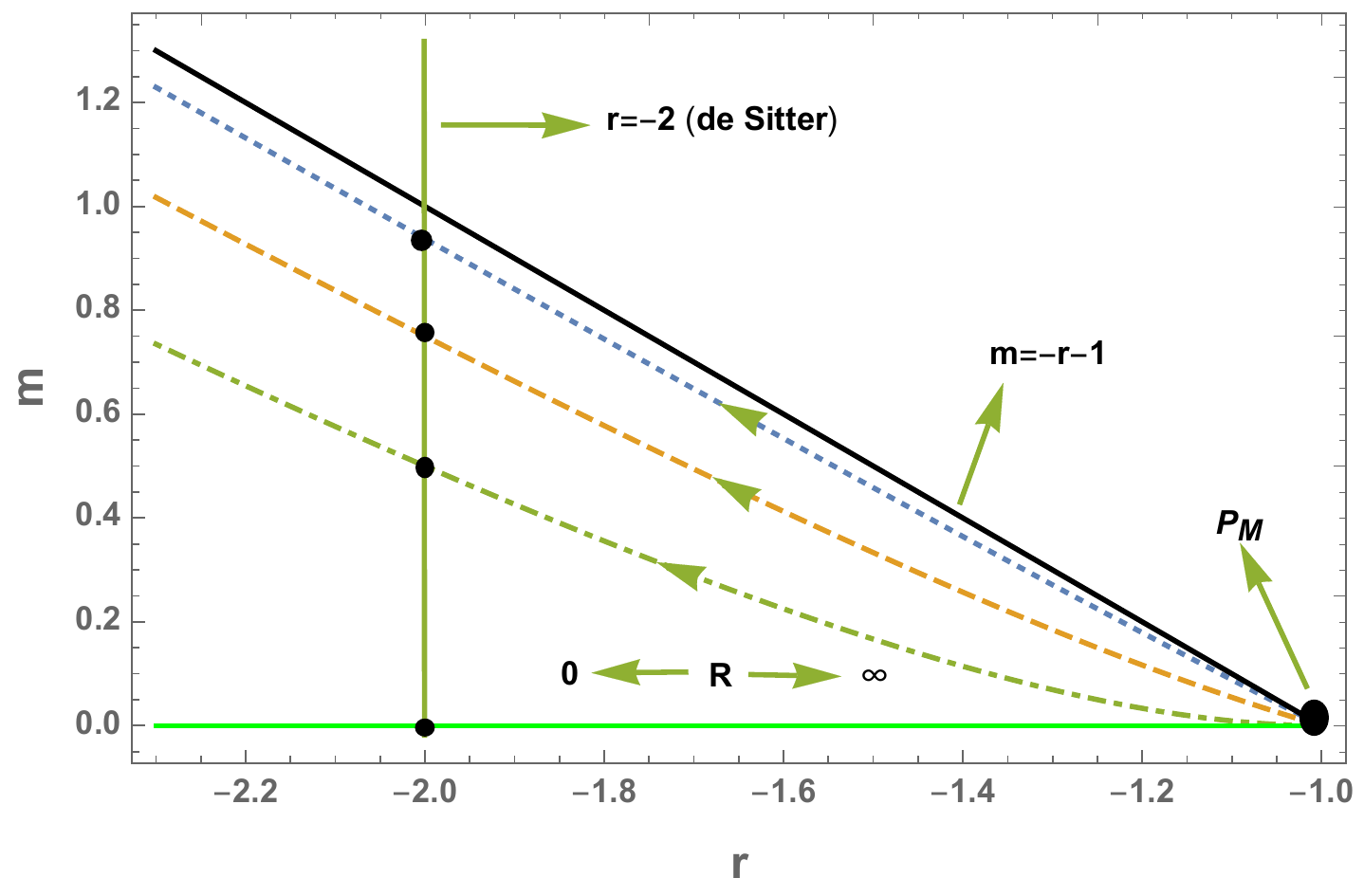}
\caption{Trajectories in the  $(r,m)$ for three different scenarios with $\eta=1/8$ (dotted), $\eta=1/2$ (Dashed) and $\eta=1$ (dot-dashed). The horizontal green line corresponds to $\Lambda$CDM, and the points correspond to the intersections with $r=-2$ which are de Sitter attractors. All trajectories connect the matter dominated saddle point $P_M$ with the late time stable de Sitter solutions at $r=-2$ with $0<m<1$.} 
\label{fig1}
\end{center}
\end{figure}
Taking the derivative of $m(r)$ at $r\rightarrow -1$ for the cases of Fig. 1 we find that for $\eta=1$, $m'(-1)=0$, which according to (\ref{autpoints}), gives large eigenvalue  $3(1+m')=3$ and the system is repelled from $P_M$ in a time shorted than the one necessary to retain the matter era 
\cite{amendola1}. For values of $\eta<1$ one finds that $-1<m'(-1)<0$, allowing the possibility of matter era as shown in the numerical example bellow. So, the power $\eta$ is fundamental in defining the viability of the model. \\
Let us check the consistency conditions $f_{,R}>0$ and $f_{,RR}>0$ with respect to the de Sitter point at $r=-2$. Using the expression for $r$ from (\ref{mr}), the de Sitter point at $R_1$ leads to 
\be\label{desitter}
\frac{R_1}{\mu^2}=\eta^{1/\eta},
\ee
and for $m(r=-2)$ we obtain
\be\label{desitter1}
m=1-\frac{\eta}{2}.
\ee
Then the stability condition of de sitter point at $R=R_1$, $0<m(r=-2)\le 1$ is satisfied by $0\le \eta <2$, which is consistent with the restriction discussed above. On he other hand, the restriction on $\eta$  given by (\ref{eta}), imply the restriction
\be\label{eta1}
0<\eta^{1/\eta}\le 1,
\ee
which leads to the following inequality between Ricci scalar at de Sitter point and the curvature scale $\mu^2$
\be\label{mudesitter}
R_1\le \mu^2
\ee 
In Fig. 2 the cosmic evolution for the model is shown in terms of the $e$-folding variable $N=\ln a=-\ln(1+z)$.
\begin{figure}
\begin{center}
\includegraphics[scale=0.7]{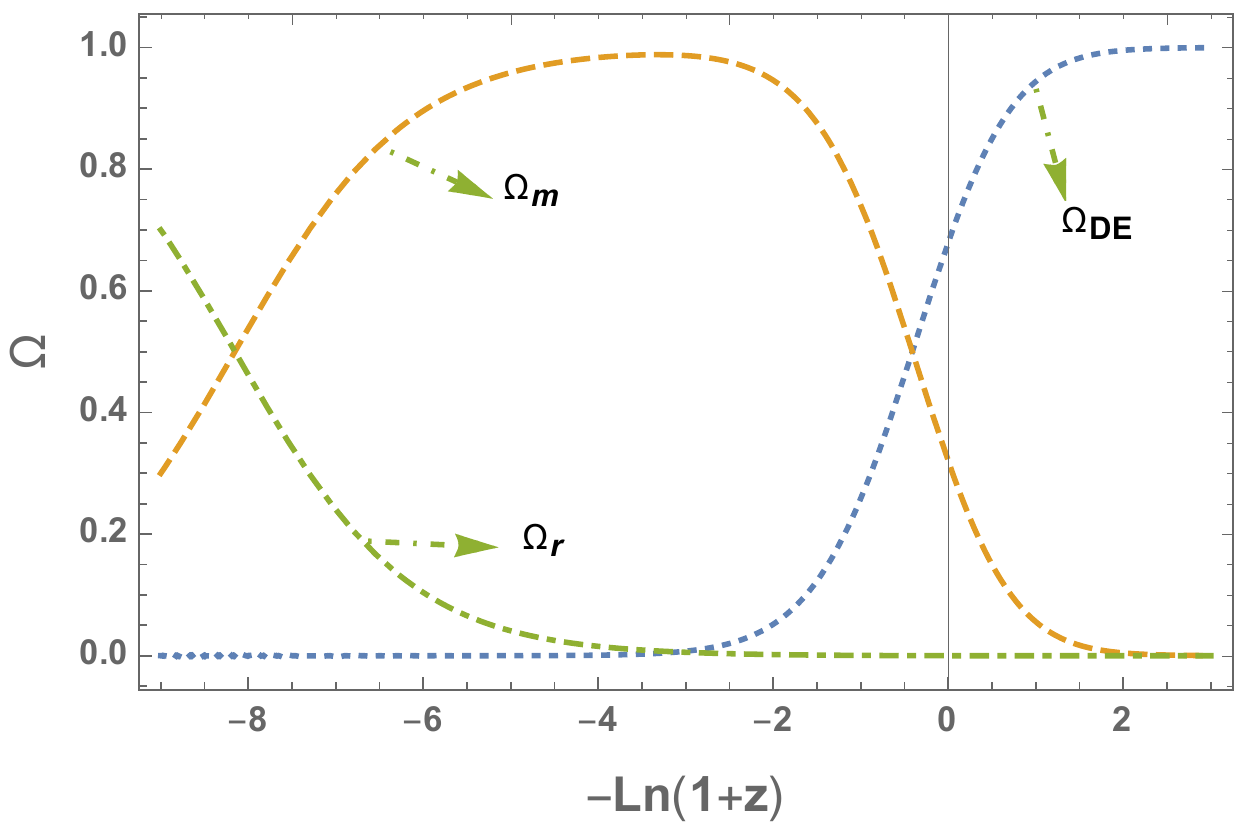}
\caption{The cosmic evolution for the matter density $\Omega_m$, dark energy density $\Omega_{DE}$ and radiation density $\Omega_{rad}$ parameters for the model (\ref{model}) with $\eta=0.68$ and initial conditions $x(-9.0)=0$, $y(-9.0)=-7.3\times 10^{-4}$, $z(-9.0)=7.300007\times 10^{-4}$ and $w(-9.0)=0.7$. The matter era lasts an adequate time and the energy fraction of the radiation at the present is $\Omega_{m0}\simeq 10^{-4}$. The model gives qualitative description of the cosmic evolution, showing the transition from matter to dark energy dominated era, the current accelerated expansion and future vacuum-dominance.} 
\label{fig2}
\end{center}
\end{figure}
The equation of state corresponding to the scenario of Fig.2 is shown in Fig. 3.
\begin{figure}
\begin{center}
\includegraphics[scale=0.7]{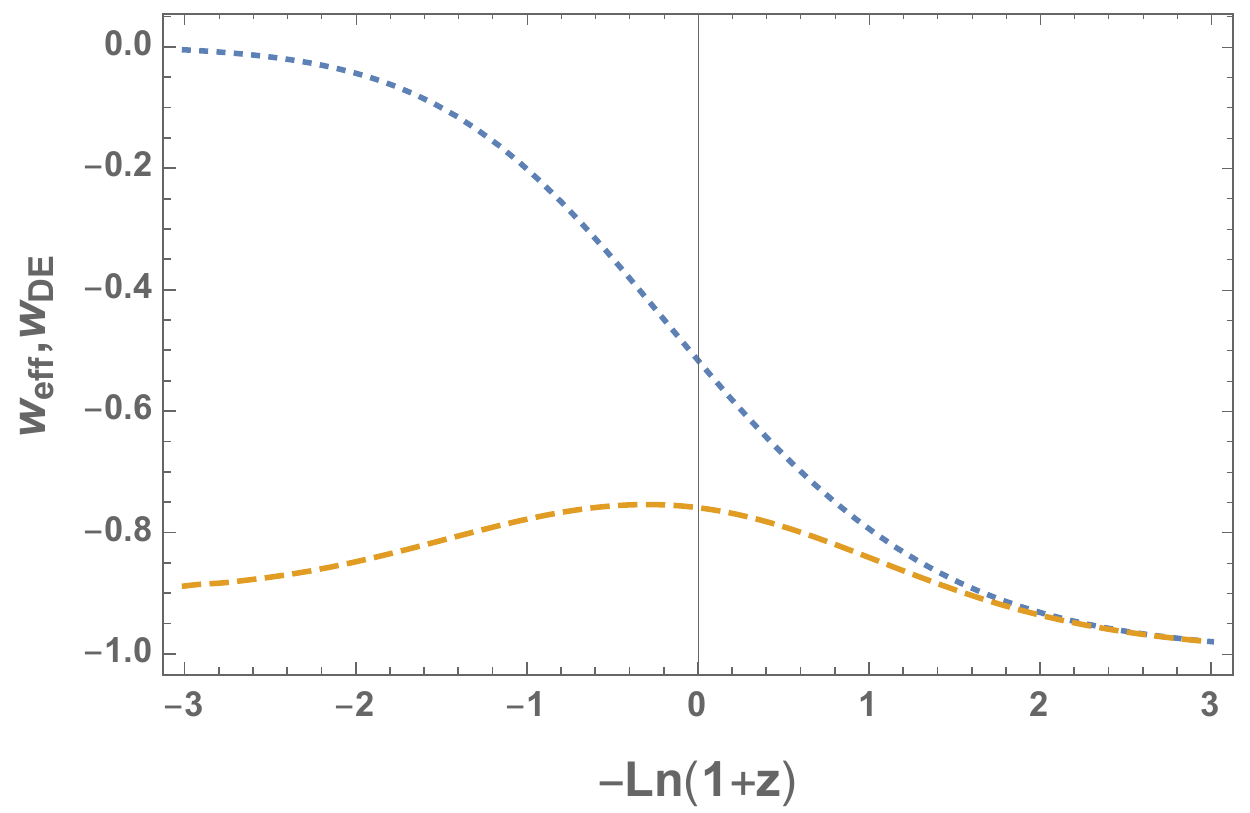}
\caption{The evolution of the effective EoS for the model (\ref{model}) with $\eta=0.68$ and initial conditions  $x(-9.0)=0$, $y(-9.0)=-7.3\times 10^{-4}$, $z(-9.0)=7.300007\times 10^{-4}$ and $w(-9.0)=0.7$. The bottom curve describes $w_{DE}$, where the approximation $F\sim F_0$ was used. Though the current $w_{DE}\sim -0.8$, the equation of state  evolves  towards the de Sitter phase.} 
\label{fig3}
\end{center}
\end{figure}
As can be seen from Figs. 2 and 3 the expansion rate is a bit slower compared to the $\Lambda$CDM model.\\
%%%%%%%%%%%%%%%%%%%%%%%%%%%%%%%%%%%%%%%%%%%%%%%%%%%%%%

\noindent {\bf Local Gravity Constraints.}\\
The effective mass of the modified gravity $f(R)$ model is given by
\be\label{effmass}
M^2=\frac{R}{3}\left(\frac{f_{,R}}{Rf_{,RR}}-1\right)=\frac{R}{3m}\left(1-m\right),
\ee
which under the condition $m<<1$, can be reduced to 
\be\label{effmass1}
M^2\simeq \frac{R}{3m}\simeq \frac{1}{3f_{,RR}}. 
\ee
The local gravity constraints are satisfied if $M\ell>>1$, where $\ell$ is the typical scale at which the gravity is measured. From (\ref{effmass1}) this constraint can be expressed in terms of $m$ as
\be\label{effmass2}
m(R_s)<<\ell^2 R_s,
\ee
where $R_s$ is the curvature of the local structure, and we assumed $f_{R_s}\simeq 1$. Making use of the relationship $R\sim H^2\sim 8\pi G\rho$ applied to the current universe ($R_0, \, \rho_0$) and to the local structure ($R_s,\, \rho_s$), we can write $R_s\sim H_0^2\rho_s/\rho_0$ and the above constraint becomes  \cite{tsujikawa1}
\be\label{effmass3}
m(R_s)<<\frac{\rho_s}{\rho_0}\left(\frac{\ell}{H_0^{-1}}\right)^2.
\ee
Applied to the current universe with $\ell\sim H_0^{-1}$, it leads to $m(R_0)<<1$. For a local structure with $\ell<<H_0^{-1}$ one expects that $m$ is even much smaller than the previous case. Thus, for the solar system with $\rho_s\sim 10^{-23}\, gr/cm^3$ and $\ell\sim 10^{13} \, cm$ one finds that $m<<10^{-24}$, where we used $H_0^{-1}\sim 10^{28} \, cm$. In order to find which value of $\mu^2$ can satisfy this restriction we use the expression (\ref{mr}) for $m$ in terms of the curvature. 
In general, for $0<\eta<1$ and $b<<1$, one has for $m<<b$ from (\ref{mr}) that
\be\label{mlocal}
\frac{\mu^2}{R}<<\left(\frac{b}{\eta}\right)^{1/\eta}.
\ee
Applied to the solar system with $b=10^{-24}$, we find
\be\label{msolar}
\mu^2<<\left(\frac{10^{-24}}{\eta}\right)^{1/\eta}10^6 H_0^2,
\ee
Taking, for instance  $\eta=0.6$, one finds that $\mu<<10^{-17}H_0$, which is much smaller than the Hubble scale today, but in terms of the $f(R)$ mass $M$ gives $M>>10^{15}H_0\sim 10^{-18} ev$, which is the expected bound. So, with an adequate choice of the parameter $\mu$ the model passes local system tests.\\
%Note that this bound is exactly the same one would expect from General Relativity if one imposes the condition of absence of additional interactions characterized by the range of interaction $M^{-1}$.\\
%%%%%%%%%%%%%%%%%%%%%%%%%%%%%%%%%%%%%%%%%%%%%%%%%%%%%%%%

\noindent {\bf Einstein Frame Potential.}\\
In the Einstein frame, in terms of the equivalent scalar field \cite{weltman1, weltman2}
\be\label{conformal}
f_{,R}=\exp\left[-{\sqrt{\frac{2}{3}}\frac{\phi}{M_p}}\right],
\ee
the potential is given by the following expression
\be\label{potentialr}
V(R(\phi))=\frac{M_p^2}{2}\frac{Rf(R)_{,R}-f(R)}{\left(f_{,R}\right)^2}.
\ee
From (\ref{model}) and (\ref{conformal}) one finds the scalar curvature as
\be\label{conformal1}
R=\mu^2\left[-\frac{1}{\eta} -W\left[-\frac{1}{\eta}e^{-\frac{1}{\eta}-\sqrt{\frac{2}{3}}\frac{\phi}{M_p}}\right]\right]^{-1/\eta},
\ee
which gives, from (\ref{potentialr}), the explicit expression for the scalar field potential 
\be\label{potentialphi}
%V=\frac{\mu^2 M_p^2}{2}\left[-\frac{1}{\eta}-W\left[-\frac{1}{\eta}e^{-\frac{1}{\eta}-\sqrt{\frac{2}{3}}\frac{\phi}{M_p}}\right]\right]^{-1/\eta}\left[e^{-\sqrt{\frac{2}{3}}\frac{\phi}{M_p}}-e^{\frac{1}{\eta}+W\left[-\frac{1}{\eta}e^{-\frac{1}{\eta}-\sqrt{\frac{2}{3}}\frac{\phi}{M_p}}\right]}\right]e^{2\sqrt{\frac{2}{3}}\frac{\phi}{M_p}}.
V=\frac{\mu^2 M_p^2}{2}\frac{\left[e^{-\sqrt{\frac{2}{3}}\frac{\phi}{M_p}}-e^{\frac{1}{\eta}+W\left[-\frac{1}{\eta}e^{-\frac{1}{\eta}-\sqrt{\frac{2}{3}}\frac{\phi}{M_p}}\right]}\right]e^{2\sqrt{\frac{2}{3}}\frac{\phi}{M_p}}}{\left[-\frac{1}{\eta}-W\left[-\frac{1}{\eta}e^{-\frac{1}{\eta}-\sqrt{\frac{2}{3}}\frac{\phi}{M_p}}\right]\right]^{1/\eta}}.
\ee
Note that the argument of the $W$-function is well defined for $0\le\phi<\infty$ ($0<\eta\le 1$).
%, but the interesting upper limit for $\phi$ is reached at the de Sitter point, where the curvature takes the value $R_1$ that satisfies (\ref{desitter}). Using (\ref{conformal}) gives the scalar field at de Sitter point as
%\be\label{phidesitter}
%\phi_{deS}=\sqrt{\frac{3}{2}}\left(\frac{1}{\eta}-\ln 2\right)M_p.
%\ee
%Replacing this value in (\ref{potential phi}) gives the potential at de Sitter point as
%\be\label{vdesitter}
%V_{deS}=\frac{1}{8}\frac{e^{1/\eta}\left(2-e^{\frac{2}{\eta}+W\left[-\frac{2}{\eta}e^{-2/\eta}\right]}\right)}{\left(-\frac{1}{\eta}-W\left[-\frac{2}{\eta}e^{-2/\eta}\right]\right)^{1/\eta}}
%\ee
The shape of the potential in the interval $0\le\phi\le2$ is shown in Fig. 4. \\
\begin{figure}
\begin{center}
\includegraphics[scale=0.7]{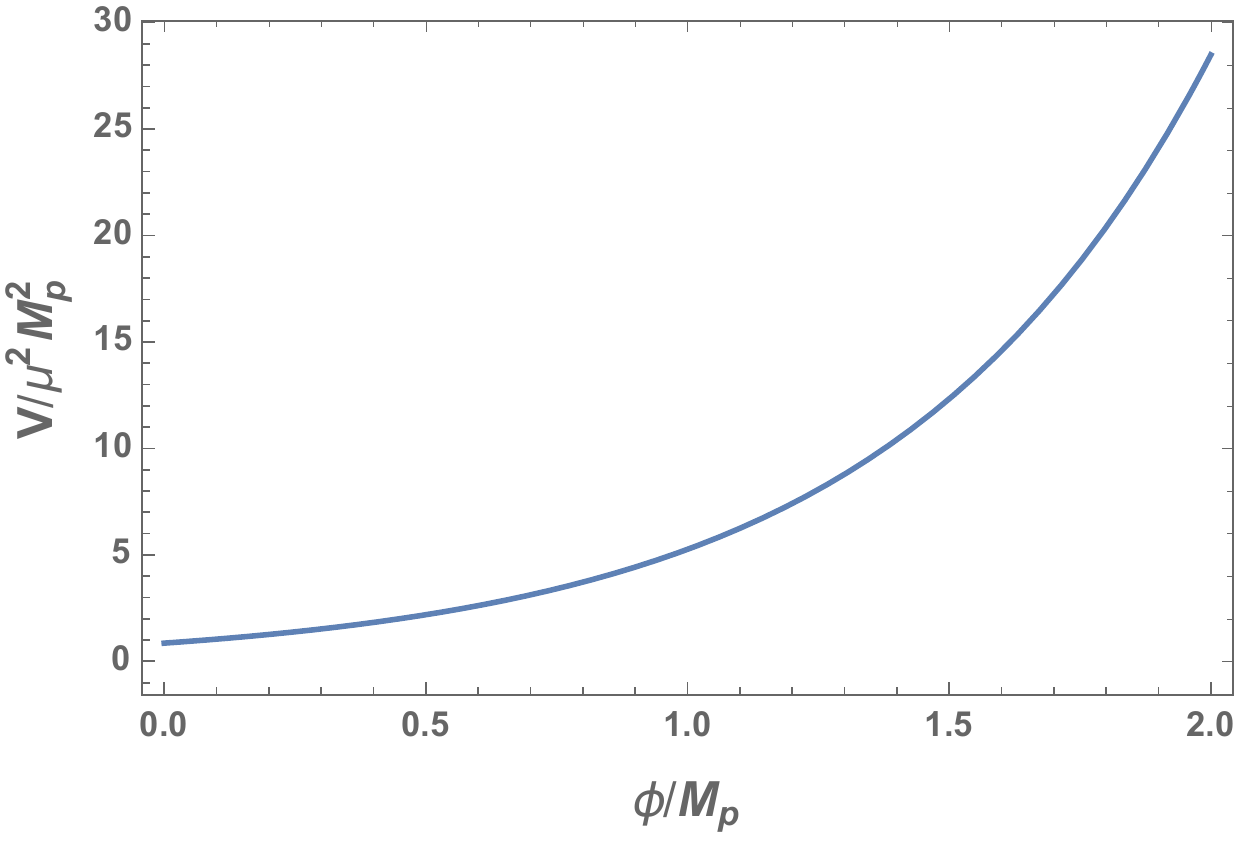}
\caption{The behavior of the potential in the Einstein frame for the model (\ref{model}) with $\eta=3/5$. The potential is represented in units of $\mu^2M_p^2$.} 
\label{fig4}
\end{center}
\end{figure}
%%%%%%%%%%%%%%%%%%%%%%%%%%%%%%%%%%%%%%%%%%%%%%%%%%%%%%%%%%%
%\{\color{red}

\noindent {\bf Model 2.}\\
A second, viable model, has the following form
\be\label{model2}
f(R)=R-2\lambda\mu^2 e^{-\left(\frac{\mu^2}{R}\right)^{\eta}}
\ee
where $\lambda$ is positive dimensionless and $\eta$ is real positive. This model satisfies the limits
\be\label{limmodel2}
\lim_{R\to \infty}f(R) =R-2\lambda\mu^2,\;\;\; \lim_{R\to 0}f(R)=0.
\ee
where the first limit leads to consistency with $\Lambda$CDM at high redshift, and the limit $R\to 0$ leads to disappearing of the cosmological constant and asymptotical flat spacetime, allowing the possibility of pure geometrical explanation of the dark energy problem. This model encodes the correction to the Einstein gravity in the form of convergent series of negative powers of curvature. The limiting case of $\Lambda$CDM can be reached not only at high curvature but also
at $\eta\rightarrow 0$ with cosmological constant $\Lambda\rightarrow e^{-1}\lambda\mu^2$. Notice that at $R\rightarrow 0$, the power $\eta$, if $\eta<<1$, has the effect of slowing the trend to zero of $e^{-\left(\mu^2/R\right)^{\eta}}$ which makes the exponential term relevant, even at current epoch, to maintain the net value of the correction to $R$ in (\ref{model2}) between the same order of magnitude over an extended cosmological period, which is important to reproduce the $\Lambda$CDM cosmology.
As in the case of the first model, taking the derivatives of (\ref{model2}) it can be seen that $f_{,R}\rightarrow 1$ and $f_{RR},\; f_{RRR},... \rightarrow 0$ at $R\rightarrow \infty$ showing that the model can avoid type I and type III singularities \cite{eelizalde}.\\
To check the viability of this model we analyze the parameters $m$ and $r$ to prove the existence of saddle matter era, i.e. $m(r\to-1^{-})>0$ and $-1<m'(r\to-1^{-})\le 0$.
\be\label{r2}
r=\frac{2\eta\lambda\mu^2\left(\frac{\mu^2}{R}\right)^{\eta}-Re^{\left(\frac{\mu^2}{R}\right)^{\eta}}}{Re^{\left(\frac{\mu^2}{R}\right)^{\eta}}-2\lambda\mu^2}
\ee
\be\label{m2}
m=\frac{2\eta\lambda\mu^2\left(\frac{\mu^2}{R}\right)^{\eta}\left[1+\eta\left(1-\left(\frac{\mu^2}{R}\right)^{\eta}\right)\right]}{e^{\left(\frac{\mu^2}{R}\right)^{\eta}} R-2\eta\lambda\mu^2\left(\frac{\mu^2}{R}\right)^{\eta}}
\ee
In order to analyze the stability conditions, $f_{,R}>0$, $f_{,RR}>0$, we first determine the value of $\lambda$ by fixing the de Sitter point $r=-2$ at $R=R_1$. From (\ref{r2}) it is found
\be\label{desitterlambda}
\lambda=\frac{R_1e^{\left(\frac{\mu^2}{R_1}\right)^{\eta}}}{2\mu^2\left[2-\eta\left(\frac{\mu^2}{R_1}\right)^{\eta}\right]}.
\ee
The restriction $\lambda>0$ can be solved by imposing
\be\label{lambdagzero}
\left(\frac{\mu^2}{R_1}\right)^{\eta}<\frac{2}{\eta}
\ee
Using the above expression for $\lambda$ in (\ref{m2}) we find the condition of stability at de Sitter point, $0<m(r=-2)\le1$, as 
\be\label{deSstability}
0<\left(\frac{\mu^2}{R_1}\right)^{\eta}\le \frac{\eta+3}{2\eta}-\frac{1}{2\eta}\sqrt{\eta^2+6\eta+1}
\ee
or
\be\label{deStability1}
\frac{\eta+1}{\eta}<\left(\frac{\mu^2}{R_1}\right)^{\eta}\le \frac{\eta+3}{2\eta}+\frac{1}{2\eta}\sqrt{\eta^2+6\eta+1},
\ee
where $\eta>0$. \\
Analyzing the condition $f_{,R}>0$ for $R>R_1$ we find, using (\ref{desitterlambda})
\be\label{stabilityfR}
f_{,R}=1+\frac{\eta\frac{R_1}{R} \left(\frac{\mu^2}{R}\right)^{\eta}e^{\left(\frac{\mu^2}{R_1}\right)^{\eta}-\left(\frac{\mu^2}{R}\right)^{\eta}}}{\eta\left(\frac{\mu^2}{R_1}\right)^{\eta}-2}>0,
\ee
which is equivalent to
\be\label{fprima}
\eta e^{-x^{\eta}}x^{\eta+1}<x_1\left(2-\eta x_1^{\eta}\right)e^{-x_1^{\eta}}
\ee
where $x=\mu^2/R$ and $x_1=\mu^2/R_1$. Due to the difficult to solve this inequality with exponentials, we first use the fact that $x<x_1$, which allows to change the above inequality by the following
\be\label{fprima1}
\eta e^{-x^{\eta}}x^{\eta}<\left(2-\eta x_1^{\eta}\right)e^{-x_1^{\eta}}.
\ee
Since $\eta x_1^{\eta}<2$ (see (\ref{lambdagzero})), then we can set 
\be\label{definningx1}
x_1^{\eta}=\frac{2}{p \eta},\;\;\; p>1
\ee
and write the inequality as
\be\label{fprima2}
e^{-x^{\eta}}x^{\eta}<\frac{1}{\eta}\left(2-\frac{2}{p}\right)e^{-2/(p \eta)}.
\ee
Note that the function $e^{-x^{\eta}}x^{\eta}$ is well defined in the real axis and has its maximum value $e^{-1}$ at $x=1$. Therefore, is enough to prove that $e^{-1}<\frac{1}{\eta}\left(2-\frac{2}{p}\right)e^{-2/(p \eta)}$. Given $0<\eta\le 1$, this can always be accomplished for $p\eta\gtrsim 1$. On the other hand, the value of $x_1$ proposed in (\ref{definningx1}) is consistent with the first inequality in (\ref{deSstability}) since
\be\label{consistency}
\frac{2}{p \eta} < \frac{\eta+3}{2\eta}-\frac{1}{2\eta}\sqrt{\eta^2+6\eta+1}
\ee
whenever  
\be\label{consistency1}
p>2+\sqrt{2},\;\;\; {\it and}\;\;\; 0<\eta\le 1.
\ee
The general stability condition $f_{,RR}>0$, using (\ref{desitterlambda}) leads to 
\be\label{stabilityfRR}
f_{,RR}=\frac{\eta e^{-x^{\eta}+x_1^{\eta}} x_1 x^{\eta+2}\left[1+\eta\left(1-x^{\eta}\right)\right]}{\mu^2\left(2-\eta x_1^{\eta}\right)}>0
\ee
The denominator is positive according to (\ref{lambdagzero}) and (\ref{definningx1}). Then, in order to satisfy this inequality 
we need to prove that $1+\eta\left(1-x^{\eta}\right)>0$, which leads to
\be\label{stabilityfRR1}
x^{\eta}<1+\frac{1}{\eta}.
\ee
Using the fact that  $x<x_1$ and assuming the expression (\ref{definningx1}) for $x_1$, this inequality can be satisfied if 
\be\label{stabilityfRR2}
p>\frac{2}{\eta+1},
\ee
which takes place as follows from the restriction (\ref{consistency1}). Hence the model satisfies the stability conditions $f_{,R}>0$ and $f_{,RR}>0$ for $R\ge R_1$.\\
Concerning the viability of the model, since $m$ cannot be expressed analytically in terms of $r$, we resort to the parametric plot of some trajectories in the $(r,m)$-plane, using (\ref{r2}) and (\ref{m2}). To this end, we use the variable $y=1/x=R/\mu^2$ with de Sitter value $y_1=1/x_1=R_1/\mu^2$ and will consider the representation for $x_1$ given by (\ref{definningx1}), i.e.  
\be\label{ydesitter}
y_1=(p\eta/2)^{1/\eta}.
\ee
The corresponding expressions for $r$ and $m$ become
\be\label{rp}
r=-\frac{2^{1+1/\eta}(p-1)ye^{1/y^{\eta}}-(p\eta)^{1+1/\eta}e^{\frac{2}{p\eta}}\frac{1}{y^{\eta}}}{2^{1+1/\eta}(p-1)ye^{1/y^{\eta}}-p(p\eta)^{1/\eta}e^{\frac{2}{p\eta}}},
\ee
\be\label{mp}
m=\frac{\eta(p\eta)^{1/\eta}e^{\frac{2}{p\eta}}\frac{1}{y^{\eta}}\left[1+\eta\left(1-\frac{1}{y^{\eta}}\right)\right]}{2^{1/\eta}\left(2-\frac{2}{p}\right)ye^{1/y^{\eta}}-\eta(p\eta)^{1/\eta}e^{\frac{2}{p\eta}}\frac{1}{y^{\eta}}},
\ee
where we used the expression for $\lambda$ from (\ref{desitterlambda}). It can be checked that, at $y=(p\eta/2)^{1/\eta}$, $r$ takes the value $r=-2$. In Fig. 5
we present some trajectories in the $(r,m)$-plane
\begin{figure}
\begin{center}
\includegraphics[scale=0.7]{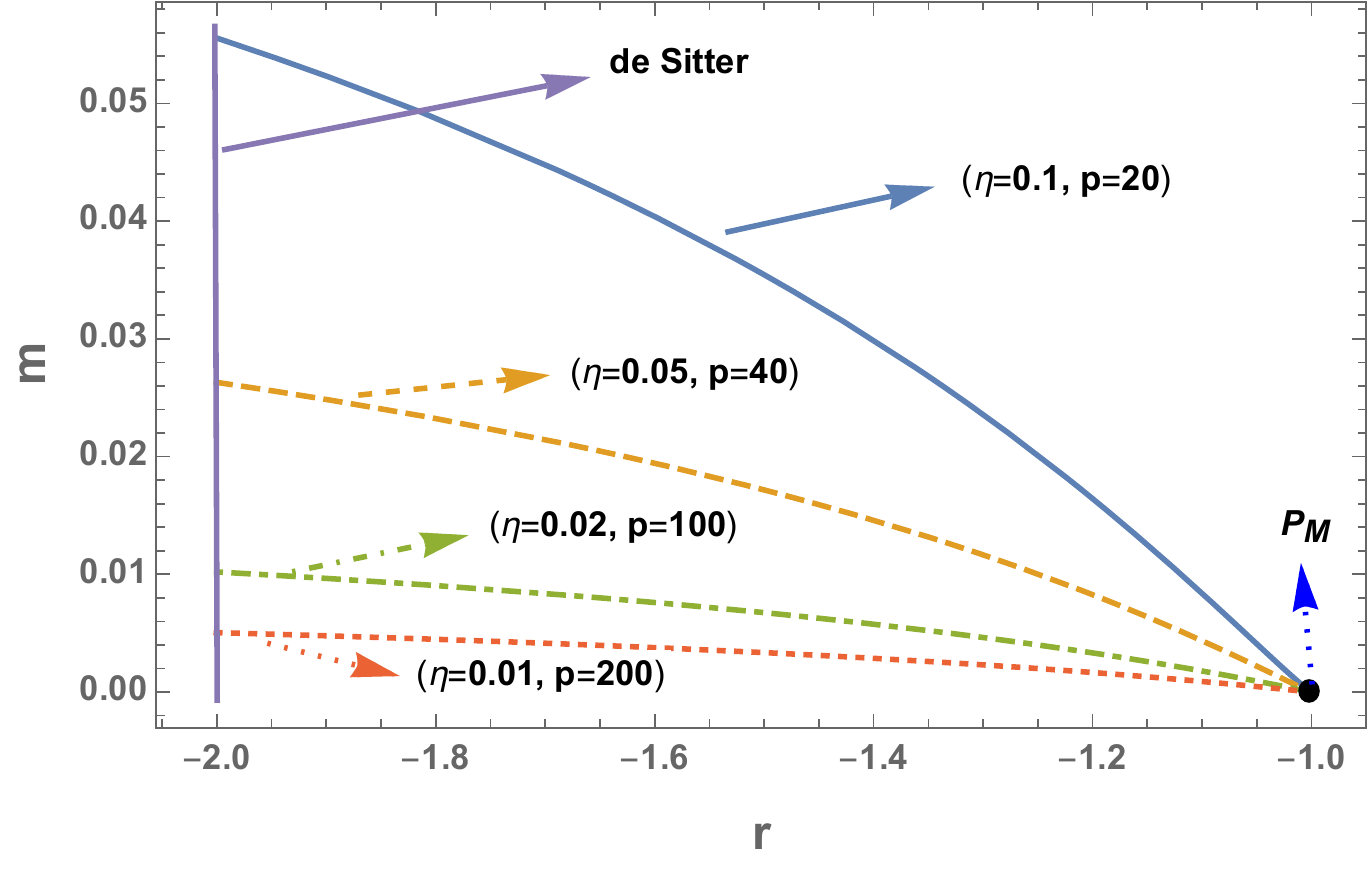}
\caption{Trajectories in the  $(r,m)$-plane for four different scenarios with $(\eta,p)=(0.1,20),\; (0.05,40),\; (0.02,100),\; (0.01,200)$. In all cases $p\eta=2$, but for smaller $\eta$ and larger $p$ the trajectories become closer to $\Lambda$CDM. All trajectories connect the matter dominated saddle point $P_M$ with the late time de Sitter attractor at $r=-2$ with $0<m<1$.} 
\label{fig5}
\end{center}
\end{figure}
The local gravity constraints can be addressed using the representation for $m$ given by (\ref{mp}). Considering, for instance, the solar system one has $y_s=R_s/\mu^2$, where $R_s\simeq 10^6 H_0^2$. As discussed before, the solar system constraints demand $m<<10^{-24}$. For the parameters $\eta$ and $p$ as used in 
Fig.5, we find that if we set $\mu^2=10^{-16}H_0^2$, then $y_s=10^{22}$ and 
\be\nonumber
\begin{aligned}
&(\eta=0.1,\; p=20)\;\;\; \Rightarrow m=9.8\times 10^{-26}\\
&(\eta=0.05,\; p=40)\;\;\; \Rightarrow m=5.3\times 10^{-25}\\
&(\eta=0.02,\; p=100)\;\;\; \Rightarrow m=7\times 10^{-25}\\
&(\eta=0.01,\; p=200)\;\;\; \Rightarrow m=4.5\times 10^{-25}
\end{aligned}
\ee
Hence, the model (\ref{model2}) can pass solar system tests, assuming $\mu\sim 10^{-8}H_0$ for the viable trajectories depicted in Fig. 5. If we consider larger values, for instance $\eta=0.5$ and $p=4$, then taking $\mu=10^{-6}H_0$ we find $m=1.3\times 10^{-27}$ and $\mu=10^{-7}H_0$ gives $m=1.3\times 10^{-30}$, which improves the results for local systems tests. \\

\noindent In order to analyze the cosmic evolution of the main density parameters $\Omega_m$, $\Omega_{DE}$ and $\Omega_r$ for the model (\ref{model2}) one needs to solve the dynamical system (\ref{restriction})-(\ref{aut4}) with appropriate initial conditions. Since there is no explicit expression for $m(r)$, we resort to an approximation, by making a polynomial fit to the paths depicted in Fig.5. Taking, for instance, the cosmological scenario with $\mu=0.01$ and $p=200$, the corresponding trajectory in Fig. 5 can be approximated by the following function of the dynamical variables $y[t]$ and $z[t]$ ($t=-\ln(1+z)$)
\be\label{m(yz)}
m=c_0+c_1\sqrt{-\frac{z[t]}{y[t]}}+c_2 \frac{z[t]}{y[t]}
\ee
with
$$ c_0=-0.0361869`,\;\; c_1=0.0533559`,\;\;  c_2=0.0171674` $$
where (`) represents more digits taken into account for the numerical calculations. In Fig. 6 we show the evolution of the main density parameters for this case.
\begin{figure}
\begin{center}
\includegraphics[scale=0.7]{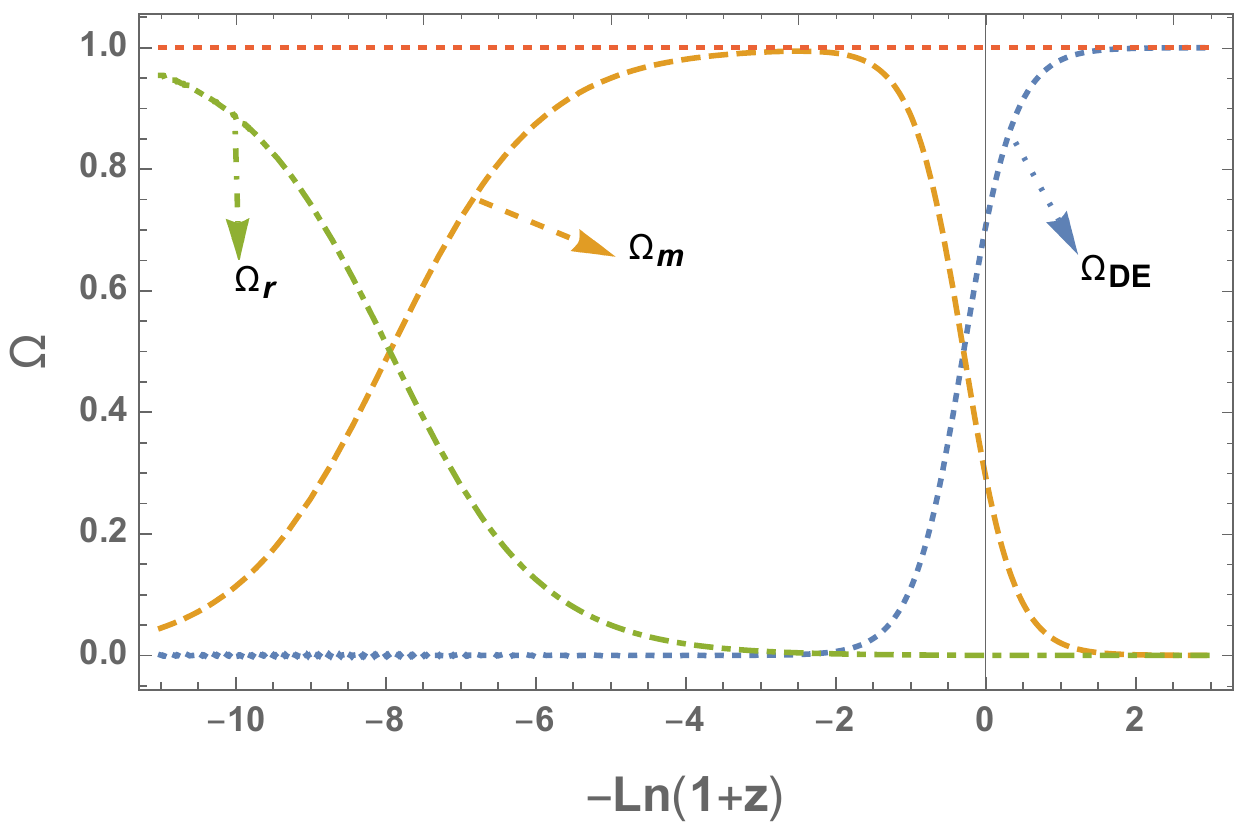}
\caption{The cosmic evolution of the density parameters for matter, radiation and dark energy for the model (\ref{model2}). In this example 
we take the path of Fig. 5 for the parameters $\eta=0.01$ and $p=200$ and used the numerical fit for $m(r)$ given by the Eq.  (\ref{m(yz)}), with initial conditions $x(-5)=0$, $y(-5)=-0.5$, $z(-5)=0.5000016$ and $w(-5)=0.05$. The behavior is compatible with the current cosmic observations on the evolution of density parameters. The obtained current densities are $\Omega_m\simeq 0.3$, $\Omega_{DE}\simeq 0.7$ and $\Omega_r\simeq 10^{-4}$.}
\label{fig6}
\end{center}
\end{figure}
The evolution of the effective and (geometry) dark energy equations of state for this numerical sample is shown in Fig. 7, where the late time similarity with the $\Lambda$CDM model is evident. \\
\begin{figure}
\begin{center}
\includegraphics[scale=0.7]{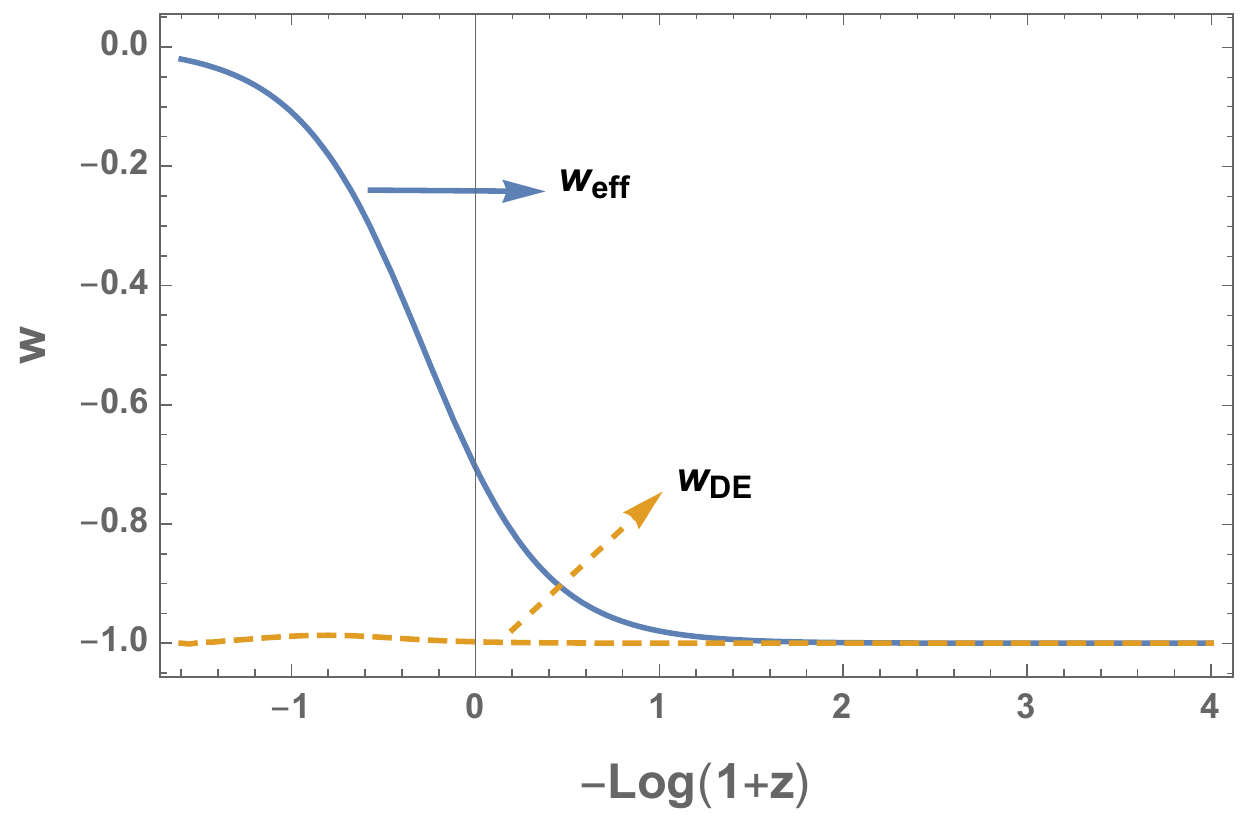}
\caption{The effective equation of state $w_{eff}$ and the equation of state associated with the geometric dark energy $w_{DE}$ for the cosmological evolution of the density parameters described in Fig. 6. The initial conditions lead to a scenario very close to the $\Lambda$CDM.}
\label{fig7}
\end{center}
\end{figure}
%%%%%%%%%%%%%%%%%%%%%%%%%%%%%%%%%%%%%%%%%%%%%%%%%%%%%%

\noindent Turning to the Einstein frame we can write the scalar field and the potential, using (\ref{conformal}) and (\ref{potentialr}), in terms of $y=R/\mu^2$ as
\be\label{phir}
\phi=-\sqrt{\frac{3}{2}}\ln \left[1-\frac{2^{-1/\eta}(\eta p)^{1+1/\eta}e^{\frac{2}{p\eta}}e^{-1/y^{\eta}}}{(2p-2)y^{\eta}}\right]
\ee
\be\label{vr}
V=\frac{2^{1/\eta}(p\eta)^{1/\eta}e^{\frac{2}{p\eta}}e^{1/y^{\eta}}(p-1)p y^2\left(1-\frac{\eta}{y^{\eta}}\right)}{\left((p-1)2^{1+1/\eta}e^{1/y^{\eta}}y-e^{\frac{2}{p\eta}}(p\eta)^{1+1/\eta}-\frac{1}{y^{\eta}}\right)^2}
\ee
where we used (\ref{desitterlambda})  and (\ref{ydesitter}). The behavior of the potential for the trajectories depicted in Fig. 5 is shown in Fig.8. It is worth noticing that the trajectories in Fig. 5 correspond to the Jordan frame, which is related to the Einstein frame by conformal transformation with conformal factor $\sqrt{f_{,R}}$ (affecting time and length scales). Therefore we can conclude that the behavior of the potential depicted in Fig. 8 shows that the parametrization we used in the Jordan frame gives also consistent results for the potential in the Einstein frame (namely, the runaway behavior of the potential which leads to the dark energy dominance at late times). Since for cosmologically viable models $f_{,R}\approx 1$, which is accomplished in our numerical case, the results in the Einstein frame are closely related to the corresponding physical magnitudes in the Jordan frame.\\ 

\begin{figure}
\begin{center}
\includegraphics[scale=0.7]{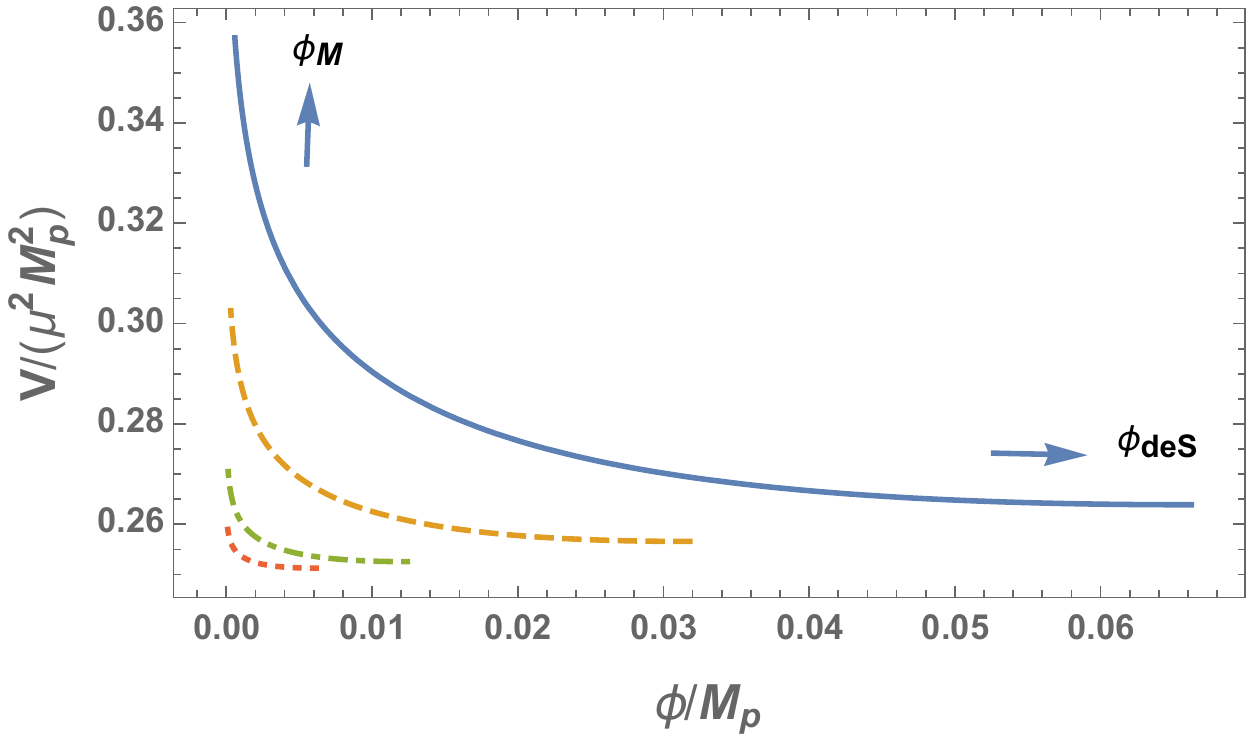}
\caption{The potential for the four scenarios with different $(\eta,p)$ depicted in Fig. 5. The arrows indicate the direction in which the matter point $P_M$ and  de Sitter point $P_{dS}$ are reached in the JF. This form of the potential shows slow-roll behavior, necessary for dark energy dominance, as the scalar field evolves towards de Sitter phase.}
\label{fig8}
\end{center}
\end{figure}

%%%%%%%%%%%%%%%%%%%%%%%%%%%%%%%%%%%%%%%%%%%%%%%%%%%%%%%%
\section{Discussion}
The role of an exponential  function of de form $\exp\left[-\left(\frac{\mu^2}{R}\right)^{\eta}\right]$ in the modified gravity is studied. Two models are proposed.
An $f(R)$ model that can satisfy cosmological and local gravity constraints is proposed. In this model the scalar curvature is multiplied by an exponential factor of the inverse curvature. The factor $e^{-(\mu^2/R)^{\eta}}$ ($\eta>0$), which tends to $1$ as $R\rightarrow \infty$, implying that $f(R)\rightarrow R$, i.e. the regular General Relativity is recovered at early times which is important to satisfy the tight constraints from Big Bang nucleosynthesis and the CMB. This is also important for the high curvature typical of local systems tests. On the other hand at  $R\rightarrow 0$, $f(R)$ satisfies the condition $f(0)=0$ i.e. the model contains the flat space-time solution without cosmological constant.  This model also satisfies the general conditions of evading    antigravity regime, $f_{,R}>0$ which is valid for any $\eta>0$, and stability, $f_{,RR}>0$ which is satisfied for $0<\eta\le 1$. This last condition implies that the flat space-time solution is stable. An important aspect of the model is that these last conditions are valid for any curvature regime, without compromising the mass (curvature) parameter $\mu^2$.  
The cosmological viability of the model follows form the analysis shown in Fig.1 in which an early mater dominant era at $m=0,r=-1$ (corresponding to a saddle point with $m'(-1)>-1$) that lasts enough to allow structure formation (see Figs. 2 and 4) and evolves towards a late time accelerated universe corresponding to a de Sitter attractor at $r=-2$. The trajectories in the $(r,m)$ plane and the cosmological evolution depicted in Figs 2 and 4 are independent of the curvature scale $\mu^2$, that can be used to satisfy the local system tests. \\
The solar system restriction for this model can be satisfied if $\mu^2<<(10^{-24}/\eta)^{1/\eta}10^6H_0^2$, giving for instance, $\mu<<10^{-17}H_0$ for $\eta=0.6$.
This result is consistent with the behavior of $r$ and $m$ at the matter dominated point $P_M$, where according to the expressions (\ref{mr}) at high redshift, when 
$\mu^2<<R$, $r$ becomes very close to -1 ($r\rightarrow -1^{-}$) and $m$ becomes very close to zero ($m\rightarrow 0^{+}$). 
We performed a numerical study of the density parameters for the cosmological scenario with  $\eta=0.68$, showing that the matter era in this model lasts enough time to allow structure formation, giving qualitatively correct description of the cosmological evolution. However, the dark energy equation of state does not reach the current expected value, close to that of the cosmological constant, at least for the proposed initial conditions.
This model can be considered as a simple toy model that satisfies viability conditions and gives qualitatively appropriate description of the cosmological evolution since the radiation and matter dominated eras.\\
The second model is more realistic since it describes the cosmological evolution consistent with the current observational data, behaving  very close to the $\Lambda$CDM. This model considers a correction to the Einstein term that disappears in the limit $R\rightarrow 0$, containing flat spacetime solution and allowing the possibility of pure geometrical explanation of the dark energy phenomenon. It gives viable trajectories in the $(r,m)$ plane that connect the matter dominated critical point at $(-1,0)$ with the de Sitter attractor at $r=-2$ and $0<m\le 1$. With the parametrization used for the de Sitter point (with parameters $p$ and $\eta$), it was shown that the conditions of stability can be satisfied whenever $p>2+\sqrt{2}$ and $0<\eta\le 1$. On the other hand, the local gravity constraints depend on three parameters, $\mu^2$, $\eta$ and $p$. For the case of the solar system it was shown that for small values of $\eta$ and large $p$ such that $p\eta\gtrsim 1$, the model can satisfy local gravity constraints with less stringent constraints on $\mu^2$ compared to the firs model. 
The model also predicts a consistent with observations evolution of the main cosmological parameters, with $w_{eff}$ showing the transition to the accelerated phase at the currently observed $z_t\sim 0.5$, and $w_{DE}\simeq -1$. The shape of the potential in the Einstein frame  favors the slow-roll behavior of the scalar field necessary for the late time dominance of dark energy.\\
\noindent A viable model has been proposed that can explain the current epoch of cosmic acceleration through purely gravitational effects and passes local system tests, eliminating the need for dark energy. The modification to the Einstein gravity is a regular function $f(R)$ satisfying the condition $\lim_{R\to 0}f(R)=0$ and approaching the limit $f(R)\rightarrow R-2\Lambda$ at high curvature, recovering the $\Lambda$CDM. This geometrical approach to the DE problem leads to results consistent with current cosmological data and could represent an appealing alternative to the $\Lambda$CDM model, where the fine tuning is replaced by adequate initial conditions. 
Further detailed analysis of local gravity constrains is needed, and it will be also interesting to study (ongoing work) the constraints on the model coming from the background and matter density perturbations. 
%%%%%%%%%%%%%%%%%%%%%%
%The future high precision observational data will place stringent constraints on the $f(R)$ model parameters , which could allow us to find patterns in the growth of large scale structures to mark difference with $\Lambda$CDM.
%The prediction for the growth of large scale structures in this model could be sufficiently large to mark difference with $\Lambda$CDM (for instance, scale-dependent pattern in the behavior of cosmological structures).

\section*{Acknowledgments}
This work was supported by Universidad del Valle under project CI 71187.

\end{document}